# Grounding Neuroscience in Behavioral Changes using Artificial Neural Networks

Grace W. Lindsay

Department of Psychology and Center for Data Science,

New York University

**Abstract**

Connecting neural activity to function is a common aim in neuroscience. How to define and conceptualize function, however, can vary. Here I focus on grounding this goal in the specific question of how a given change in behavior is produced by a change in neural circuits or activity. Artificial neural network models offer a particularly fruitful format for tackling such questions because they use neural mechanisms to perform complex transformations and produce appropriate behavior. Therefore, they can be a means of causally testing the extent to which a neural change can be responsible for an experimentally observed behavioral change. Furthermore, because the field of interpretability in artificial intelligence has similar aims, neuroscientists can look to interpretability methods for new ways of identifying neural features that drive performance and behaviors.

**Introduction and Motivation**

A goal present in many lines of neuroscience research is to connect neural mechanisms to higher level functions, cognition, and/or behavior. While the goal is common, approaches to achieving it are varied.

Bottom-up methods study properties of neurons and neural circuits in the hopes of identifying 'canonical' computations implemented by the neural hardware [1,45]. Reverse engineering function from structure in a vacuum, however, has proven difficult, with little agreement on what important computations are implemented in common structures. The cognitive approach tries to identify intermediate mental capacities such as attention, memory, and executive control, that can guide the search for neural mechanisms across a variety of behaviors. This approach, however, faces a difficult grounding problem. To determine which behaviors rely on a cognitive capacity, scientists often look to the underlying neural mechanisms, but the neural mechanisms themselves are what neuroscience aims to identify for a given capacity [2].

Instead of neural hardware or cognitive capacities, focusing on behavior as the explicandum can ground neuroscience studies in very concrete questions such as "which features of neural activity lead to the produced behavior?" This sidesteps the need to categorize intermediate cognitive concepts and provides criteria for determining if a given neural mechanism is important.

An even more direct approach to understanding the neural mechanisms crucial for a behavior can come from identifying circumstances under which behavior changes. By studying settings wherein the same input to the same individual leads to different outputs, we can seek out the precise neural changes that drive such behavioral changes, and therefore may underlie the production of that behavior more generally. Such settings can be found in studies of development, learning, attention, internal state or arousal shifts, error trials, in-trial dynamics, adaptation, disease, history effects, or aging (see Box).

Experimental studies of this form help identify possible neural mechanisms responsible for behavioral changes.  For example, recent studies have shown how a network of cortical and subcortical areas contributes to changes in maternal behavior [28], how the hypothalamus is involved in motor adaptation [29], and how identified subpopulations of prefrontal cortex are associated with performance in an attention task [30]. When possible, findings gleaned from these studies are verified with manipulations of neural activity as well. Here I argue that artificial neural networks have an important role to play in complementing and expanding the study of how neural changes lead to behavior changes.

**Use of artificial neural networks to connect neural activity and behavior**

Artificial neural networks (ANNs) offer a productive modeling framework to tackle these questions because, unlike more traditional neural circuit models, they are explicitly built to produce behavior [3]. In this way, they can link features of neural populations to behavioral outputs.

A research trend in computational neuroscience for the past decade has been to train ANNs on large amounts of high-dimensional data to perform complex tasks, and compare the representations learned in these models to brain areas believed to be involved in the same behavior [4]. This has established several strong connections between these models and biological processing, though the exact mechanisms by which these models replicate neural activity and behavior often remain unclear.

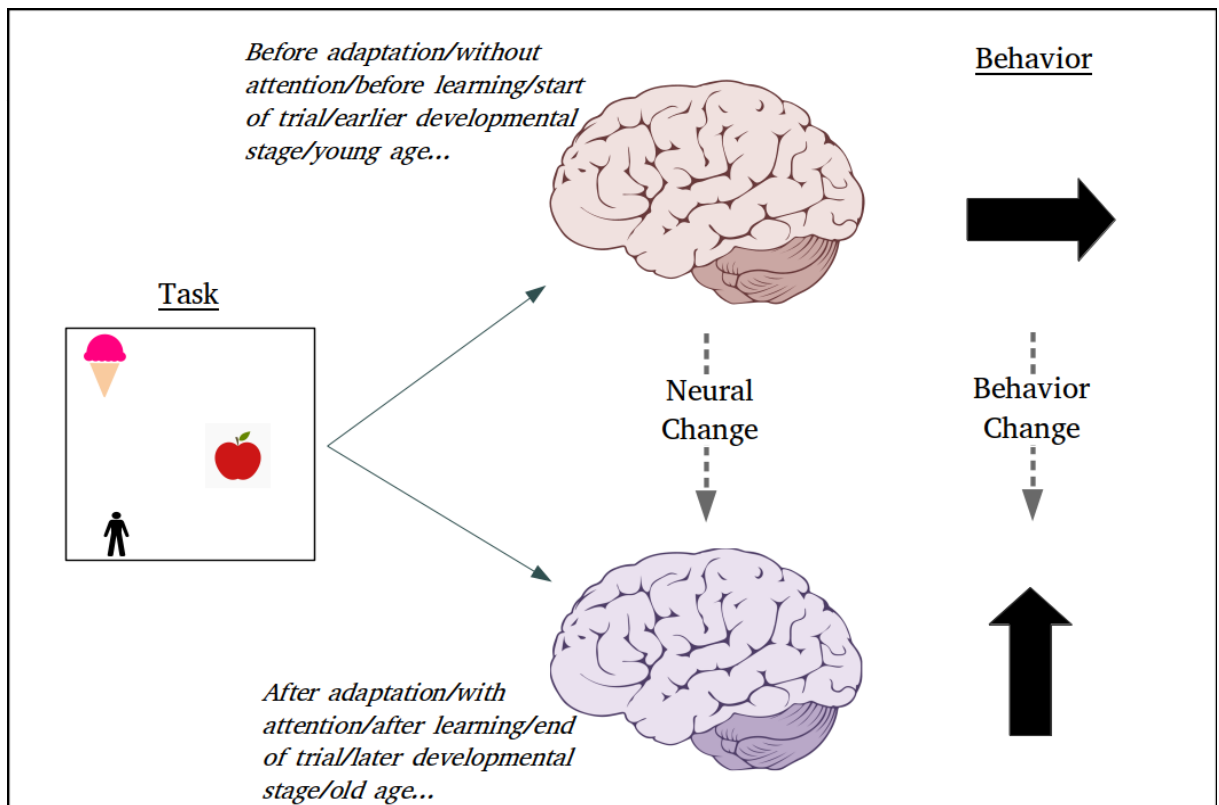

**Focusing on behavioral differences observed within an individual facing the same task under different circumstances can help target the search for the neural mechanisms that drive behavior.** Many tasks, such as decision-making, navigation, foraging (as shown here), perceptual reports, classification, detection, and motor production, provide useful behavioral readouts that can be compared across conditions. If a neural system is given the same input, changes in behavioral output must be supported by changes in neural circuitry and/or activity. Therefore, conditions under which behavior changes are ideal for identifying the neural mechanisms most directly relevant for implementing the input-output transformations that support behavior.

The approach offered here differs from much of the previous work in ANN modeling of brains because it does not necessarily aim to provide exact quantitative matches to neural activity or behavior. Rather, it aims to capture trends in performance changes under different circumstances and a description of the neural changes that cause them. In this way, it has the possibility to offer interpretable results that could also generalize across different datasets or brain regions.

Several recent studies, reviewed in the following section, have made progress on explaining the neural features underlying behavior by modeling such relations in ANNs. These ANNs perform tasks inspired by the neuroscience and psychology literature and/or use mechanisms derived from neurophysiology. The field of interpretable AI, which aims to understand the workings of ANNs in general (not just brain-inspired ones), has also identified several methods that can find the neural features responsible for network outputs. In the fourth section, recent examples of such methods from interpretable AI are reviewed.

**Connecting neural mechanisms to behavioral changes in brain-inspired ANNs**

Learning represents a change in response to stimuli after exposure. Advances in deep reinforcement learning (RL) make it possible to train neural network-based agents in settings similar to how animal subjects are trained. In an extensive study of sensorimotor learning, the authors of [9] use reinforcement learning to train thousands of artificial agents on an object manipulation task. They found a correlation between space and velocity tuning in the neural representation and performance on the task. They also demonstrated the importance of a sparse neural code and abundant state-value and policy representing neurons and a sparse neural code. By silencing these state-value and policy representing neurons, they were able to demonstrate their causal role in model performance. Impressively, the authors were able to carry out a suite of neural recording and perturbation experiments in mice verifying the emergence and role of these representations in the mouse brain.

Another reinforcement learning study from the same lab targeted learning of behaviors that are composites of previously-learned behaviors [10]. It showed that mice could learn a complex task that was a composite of previously-learned behaviors within one training session, demonstrating few-shot learning. Such learning was replicated in an RL-trained ANN. Neurons encoding relevant subtask information were identified in the ANN and their ablation demonstrated their causal role in producing the composite behavior. Specifically, ablating Q-value encoding neurons slowed learning of the composite task, whereas a control ablation of non-specific neurons did not. This points to the importance of identifying the neural representation of RL concepts when studying task learning.

A recent paper tackled performance changes in object discrimination and recognition that occur during two different learning settings: 'life-long' and 'real-time' [5]. The former is represented by ego-centric visual data collected at different developmental stages. The latter is derived from a laboratory experiment that varied unsupervised exposure to stimuli [6]. In the current study, the authors found that, while not perfect, certain 'self-supervised' algorithms for updating weights in ANNs better captured the broad trends in performance changes observed in humans. An important feature of the algorithms that best matched the data was the ability to learn unique representations in spite of the large amount of similarity in the real-world input data. They achieved this by pulling 'negative samples' from memory, pointing to a potential role of memory in perceptual learning.

Another study approached the problem of continual learning by incorporating replay [7]. Replay of past experience is believed to support efficient learning in the brain [8]. By putting a biologically-inspired replay mechanism into an ANN, this work was able to capture the ability of the brain to learn new skills without forgetting old ones. Importantly, the model did not require a large amount of replayed samples in order to prevent forgetting, making it a plausible mechanism for biological learning. This work therefore highlights the role of ANNs in verifying hypotheses about the mechanisms of learning in the brain.

In addition to task learning, sensory systems can also adapt rapidly to changing settings. In [11], the authors built an ANN model of how neurons adapt to changing background noise in a way that supports noise-invariant speech recognition. They identified gain changes and receptive field shifts in the model that supported its adaptive abilities. The authors discuss how this modeling work both supported existing ideas on auditory processing and proposed potential new mechanisms, especially regarding the dynamics of the receptive field.

Sensory systems also show rich short-timescale dynamics in response to static inputs as a result of recurrent connections. Such connections are easy to include and study in ANN models. In [12], different anatomical and functional forms of recurrence were all able to capture trends in the dynamics of human object recognition performance, though through different underlying neural dynamics. An analysis of fMRI data, suggested predictive feedback recurrence may be the neural mechanism at play in human vision.

A different study pinpointed the way recurrent connections target the representation of auxiliary variables and how this leads to improved object recognition performance over time [13]. Specifically, they showed that category-orthogonal information such as location increases over time as a result of recurrent influence and, through a clever perturbation method, they provided evidence that this information is crucial for object recognition performance (see Figure 1). By focusing on information representation at the population level, this work demonstrates how ANNs can generate hypotheses that are testable in real brains even if the low-level details of the model and brain do not align.

To identify the neural architecture behind oddity detection, the authors of [14] use an ANN model that has been shown to align well with the ventral visual stream. They show that this model predicts concurrent visual discrimination behavior across a variety of tasks for subjects with lesions of perirhinal cortex (PRC), but it does not predict performance of those with intact PRC. Through this the authors provide support for the role of PRC in producing this behavior.

Studying behavior with and without attention also provides a good opportunity for identifying the mechanisms important for perceptual performance [35]. Two recent studies have used CNNs to show how changes in neural gain associated with spatial attention positively influence performance on visual tasks. In the first, enhancement of gain at the attended region is compared to decreased gain at all other locations and both are shown to substantially increase performance on a detection task [36]. The second paper isolates the impacts of gain compared to another commonly observed neural correlate of spatial attention: receptive field changes. They found that gain changes alone are "both necessary and sufficient" to explain performance changes [37].

In total, these studies show that the neural mechanisms that drive behavior change under conditions such as adaptation, learning, and attention can be fruitful studied in ANNs. These studies can both verify existing hypotheses and identify new ones.

**Interpretable AI studies examining internal mechanisms responsible for behavior**

As demonstrated above, using ANNs to build models directly inspired by problems and data in neuroscience is a good way to test hypotheses about how behaviors are produced by neural mechanisms. The field of AI, while not directly aimed at understanding the brain, also has an interest in connecting features of ANNs to their behavior. Methods in the fields of explainable and interpretable AI, in particular, can be used to identify the neural features in trained networks that are responsible for their performance [15,16,41]. Examples of controlling network behavior by controlling neural features in interpretable AI can serve as inspiration for neuroscience; these examples provide insight into what methods are most successful at making the neuron-behavior connection and what concepts are most fruitful for thinking about this relationship. Reviewed here are recent successes organized from the low (single neuron) level to the high, with an emphasis on the methodology used to achieve behavioral control.

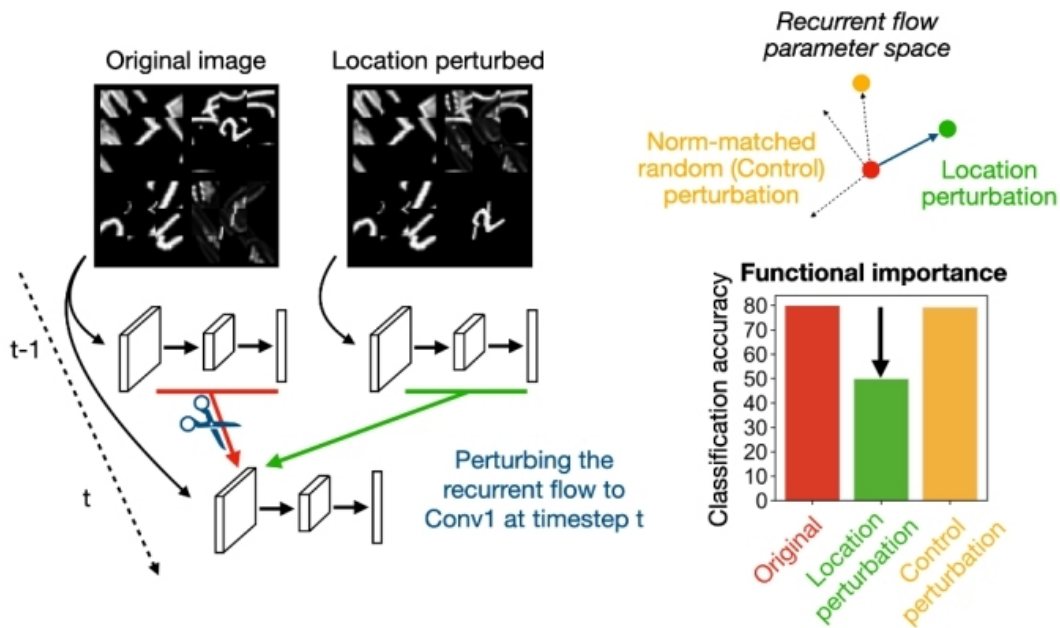

**Figure 1.** *Procedure for determining the causal relevance of recurrent influences on classification behavior. The importance of information about auxiliary variables, such as location, is determined by cutting off the recurrent influence normally elicited by an image with an object at a given location and replacing it with recurrent influence from an image with the object at a different location. If the presence of this category-orthogonal information in the recurrent influence on feedforward activity is relevant for performance, such a perturbation should lead to a decrease in classification accuracy compared to a control perturbation. Adapted from [13].*

In [42], the authors aimed to identify which neurons in a large language model (LLM) are responsible for retrieval of specific knowledge. To do this, they calculated the impact of a neuron's activity on the model output as its activity level varied, through a method known as integrated gradients. By looking at several conditions where the model produced the same factual output (e.g. responding "Dublin" to prompts like "The capital of Ireland is" and "The largest city in Ireland is"), they identifed neurons that represent knowledge of a specific concept. They showed that by changing the activity of as little as four neurons, they could control the expression of specific knowledge.

The authors of [20] also tackled the question of where LLMs store their facts. To do so they used causal mediation analysis. This method identifies important neurons in the model as those whose activity restores accurate output of the model output when the model is given corrupted inputs. Through this they were able to suggest a procedure for how facts are recalled that involves an encoding of the input concept that is passed to neurons that produce memorized properties of it and these properties go on to be integrated and summed downstream in order to produce the model output.

Looking at how populations of neurons collectively contribute to model behavior, the authors of [19] train a language model on lists of moves from the board game Othello. To determine if the model learns an internal representation of board states, the authors train a series of classifiers to read out board state from the population activity at different layers. Importantly, they find that linear classifiers (which are commonly used in neuroscience [44]) perform poorly, but nonlinear classifiers are better able to read out board state. By changing activity in a way that changes the nonlinear classifier's estimate of board state, they were able to change the moves that the model outputs. Similarly, in [17], the authors studied the neural representations of a model trained to generate images from text. By learning carefully-crafted linear readouts of the internal representations, they were able to identify orthogonal subspaces for the representation of images that contain written words and those that don't. By manipulating the network according to these subspaces, they could control whether or not a generated image would contain written words (see Figure 2).

In [18], the authors worked with models trained to classify images and identified how the classification rules are embedded in the network by focusing on the role of specific *weights.* By looking at the activity of a "pre-synaptic" set of neurons in response to one image and the activity of the "post-synaptic" population in response to an altered version of the image (wherein, e.g., a key feature of the object was changed), they could calculate how connections between these populations should change in order to change how the network relates concepts to each other. This ability to rewire connections between concepts also revealed what features the network was relying on for its classification behavior: co-occurring visual items, such as roads in images of race cars, were frequently identified as important for classification.

Another approach to identifying neural mechanisms relevant for performance is to track the neural changes that occur during periods of dramatic performance changes during model training. In this vein, the authors of [21] tried to shed light on the phenomena of 'in-context learning': the ability of an LLM to learn to mimic a style or pattern without updating its weights. By exploring a period of strong change in performance during model training, they identified the emergence of an important circuit element that implements pattern completion. By 'ablating' these circuits, they were able to impair in-context learning. Relatedly, in a study of visual processing [43], the authors noted an increase in low-level orientation selectivity that co-occured with an increase in generalization performance. The causal role of this selectivity for model performance was verified through ablation experiments.

**Limitations and Future Directions**

These recent works, coming from both the fields of computational neuroscience and AI interpretability, show that it is possible to identify neural mechanisms that are causally responsible for behavior in complex and highly-distributed neural systems. Proving causality in ANN models is much easier than attempting to do so in real neural systems, due to the ease with which these networks can be perturbed and lesioned. Given the ability of these networks to perform tasks similar to those tested in laboratories, they offer many opportunities to inspire and test hypotheses about the neural mechanisms underlying behavior.

It is important to note that 'mechanism' can take many different meanings in the neuroscience literature. Because ANNs rarely include fine-grained details of cell physiology or one-to-one correspondence with real neurons, they are unable to provide insights on that level. ANN models therefore hold the most promise for elucidating mechanisms defined closer to the 'algorithmic' level, as defined by Marr [34]  That is, they can be models of how information is represented, distributed, and transformed across neural populations and how such transformations intelligently map inputs to outputs. For models with established and intentional ties to neural anatomy, it may be possible to map these representations and transformations onto specific brain areas. But in more general models, especially in the kind of work from AI discussed above, the contribution to neuroscience will be more in the form of identifying the kinds of representations and transformations that may be helpful for a task, rather than identifying their specific implementation in the brain.

**Figure 2.** *By identifying where written word information was stored in an image-generating model, the method from [17] can control the extent that generated images contain words.*

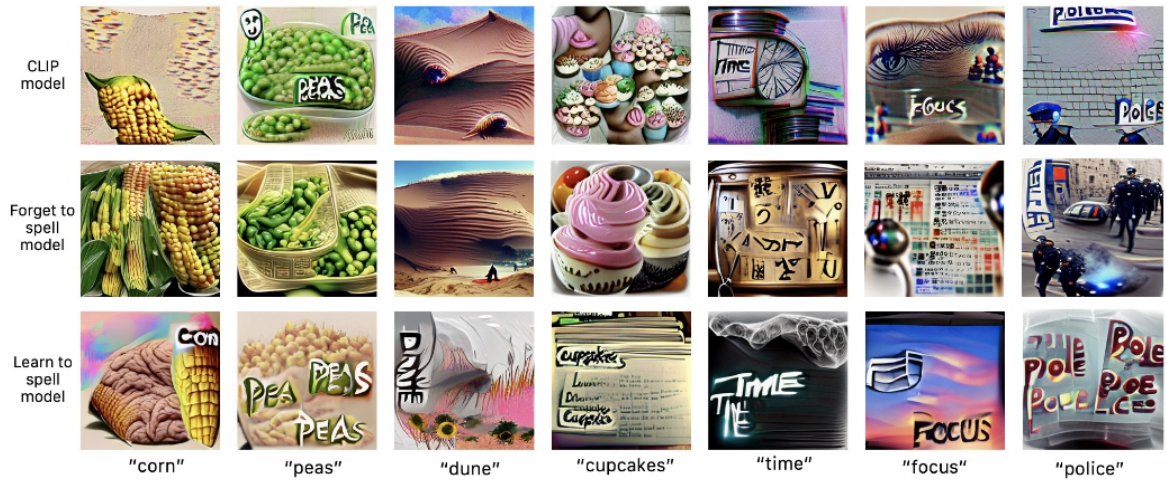

In the end, however, the mechanisms proven or discovered in these models will only provide indirect support for their role in the brain. Advances in the ability to causally manipulate real neural circuits will remain crucial for the ability to connect neural mechanisms to behavior [32,33].

A growing interest in ethology in neuroscience combined with methods that make it possible to both track behavior and record neural activity under a variety of experimental setups [22, 23]. In addition to helping connect neural activity to behavior, this has also resulted in a push for experiments that target more naturalistic behaviors. As shown above, ANNs make it possible to explore neural mechanisms behind behaviors that can involve high-dimensional inputs and complex and dynamic outputs.

While these models offer a large step-up over the simpler hand-built models that came before, they do still lack a lot of the features that may be relevant for understanding behavior of an organism in its natural environment. Past work has established frameworks for creating embodied artificial agents with multi-modal sensory inputs and species-specific architectures and actuators [24, 25, 26]. Interest in this line of work is rapidly expanding at the intersection of neuroscience and AI [27]. The combination of advanced experimental methods with models that can explore simulated environments in simulated bodies can unleash a new wave of research that tackles previously intractable questions about the neural basis of natural behavior.

## Acknowledgements

Thanks to sci-draw.io for images used in Box 1 and to the three anonymous reviewers for substantially improving this work.

## References

[1] Marcus G, Marblestone A, Dean T. The atoms of neural computation. Science. 2014 Oct 31;346(6209):551-2.

[2] Francken JC, Slors M, Craver CF. Cognitive ontology and the search for neural mechanisms: three foundational problems. Synthese. 2022 Sep 7;200(5):378. **This paper lays out three problems at the heart of cognitive neuroscience. It covers difficulties in determining which tasks rely on which cognitive capacities and in identifying which neural mechanisms matter for a cognitive capacity.**
[3] Yang GR, Wang XJ. Artificial neural networks for neuroscientists: a primer. Neuron. 2020 Sep 23;107(6):1048-70.
[4] Doerig A, Sommers R, Seeliger K, Richards B, Ismael J, Lindsay G, Kording K, Konkle T, Van Gerven MA, Kriegeskorte N, Kietzmann TC. The neuroconnectionist research programme. arXiv preprint arXiv:2209.03718. 2022 Sep 8.
[5] Zhuang C, Xiang V, Bai Y, Jia X, Turk-Browne N, Norman K, DiCarlo JJ, Yamins DL. How Well Do Unsupervised Learning Algorithms Model Human Real-time and Life-long Learning?. In Thirty-sixth Conference on Neural Information Processing Systems Datasets and Benchmarks Track 2022. *** This paper establishes benchmarks for models aiming to replicate both real-time and life-long learning. The life-long learning benchmark, in particular, is a substantial contribution as it relies on a dataset of first-person visual experience obtained in real world settings.**
[6] Jia X, Hong H, DiCarlo JJ. Unsupervised changes in core object recognition behavior are predicted by neural plasticity in inferior temporal cortex. elife. 2021 Jun 11;10:e60830.
[7] Van de Ven GM, Siegelmann HT, Tolias AS. Brain-inspired replay for continual learning with artificial neural networks. Nature communications. 2020 Aug 13;11(1):4069.
[8] Liu Y, Mattar MG, Behrens TE, Daw ND, Dolan RJ. Experience replay is associated with efficient nonlocal learning. Science. 2021 May 21;372(6544):eabf1357.
[9] Suhaimi A, Lim AW, Chia XW, Li C, Makino H. Representation learning in the artificial and biological neural networks underlying sensorimotor integration. Science Advances. 2022 Jun 3;8(22):eabn0984. *** This work offers an extensive comparison and analysis of both mice and artificial neural networks trained through reinforcement learning on the same object manipulation task.**
[10] Makino H. Arithmetic value representation for hierarchical behavior composition. Nature Neuroscience. 2022 Dec 22:1-0.
[11] Mischler G, Keshishian M, Bickel S, Mehta AD, Mesgarani N. Deep neural networks effectively model neural adaptation to changing background noise and suggest nonlinear noise filtering methods in auditory cortex. NeuroImage. 2023 Feb 1;266:119819.
[12] Lindsay GW, Mrsic-Flogel TD, Sahani M. Bio-inspired neural networks implement different recurrent visual processing strategies than task-trained ones do. bioRxiv. 2022 Mar 8:2022-03.
[13] Thorat S, Aldegheri G, Kietzmann TC. Category-orthogonal object features guide information processing in recurrent neural networks trained for object categorization. arXiv preprint arXiv:2111.07898. 2021 Nov 15. *** This work shows a counter-intuitive benefit of category-irrelevant information and how it unfolds in the neural representation as a result of recurrent connections.**
[14] Bonnen T, Yamins DL, Wagner AD. When the ventral visual stream is not enough: A deep learning account of medial temporal lobe involvement in perception. Neuron. 2021 Sep 1;109(17):2755-66.
[15] Linardatos P, Papastefanopoulos V, Kotsiantis S. Explainable ai: A review of machine learning interpretability methods. Entropy. 2020 Dec 25;23(1):18.

[16] Gilpin LH, Bau D, Yuan BZ, Bajwa A, Specter M, Kagal L. Explaining explanations: An overview of interpretability of machine learning. In2018 IEEE 5th International Conference on data science and advanced analytics (DSAA) 2018 Oct 1 (pp. 80-89). IEEE.
[17] Materzyńska J, Torralba A, Bau D. Disentangling visual and written concepts in CLIP. InProceedings of the IEEE/CVF Conference on Computer Vision and Pattern Recognition 2022 (pp. 16410-16419).
[18] Santurkar S, Tsipras D, Elango M, Bau D, Torralba A, Madry A. Editing a classifier by rewriting its prediction rules. Advances in Neural Information Processing Systems. 2021 Dec 6;34:23359-73.
[19] Li K, Hopkins AK, Bau D, Viégas F, Pfister H, Wattenberg M. Emergent world representations: Exploring a sequence model trained on a synthetic task. arXiv preprint arXiv:2210.13382. 2022 Oct 24. **\* This paper is a clearly laid out project that uses a simple board game to causally probe the mechanisms of language models.**
[20] Meng K, Bau D, Andonian AJ, Belinkov Y. Locating and editing factual associations in gpt. InAdvances in Neural Information Processing Systems 2022 Feb 10.
[21] Olsson C, Elhage N, Nanda N, Joseph N, DasSarma N, Henighan T, Mann B, Askell A, Bai Y, Chen A, Conerly T. In-context learning and induction heads. arXiv preprint arXiv:2209.11895. 2022 Sep 24.
[22] Pereira TD, Shaevitz JW, Murthy M. Quantifying behavior to understand the brain. Nature neuroscience. 2020 Dec;23(12):1537-49.
[23] Nastase SA, Goldstein A, Hasson U. Keep it real: rethinking the primacy of experimental control in cognitive neuroscience. NeuroImage. 2020 Nov 15;222:117254. **\*\* This perspective warns against overuse of simplistic laboratory experiments in cognitive neuroscience. Specifically it argues that the rules and principles learned in such controlled settings do not necessarily generalize to natural, evolutionarily-determined behaviors and it lays out ideas for approaching ecologically-grounded experiments.**
[24] Bhattasali NX, Zador AM, Engel TA. Neural circuit architectural priors for embodied control. arXiv preprint arXiv:2201.05242. 2022 Jan 13.
[25] Merel J, Aldarondo D, Marshall J, Tassa Y, Wayne G, Ölveczky B. Deep neuroethology of a virtual rodent. arXiv preprint arXiv:1911.09451. 2019 Nov 21.
[26] Lobato-Rios V, Ramalingasetty ST, Özdil PG, Arreguit J, Ijspeert AJ, Ramdya P. NeuroMechFly, a neuromechanical model of adult Drosophila melanogaster. Nature Methods. 2022 May;19(5):620-7.
[27] Zador A, Richards B, Ölveczky B, Escola S, Bengio Y, Boahen K, Botvinick M, Chklovskii D, Churchland A, Clopath C, DiCarlo J. Toward next-generation artificial intelligence: catalyzing the NeuroAI revolution. arXiv preprint arXiv:2210.08340. 2022 Oct 15.
[28] Tasaka GI, Hagihara M, Irie S, Kobayashi H, Inada K, Kihara M, Abe T, Miyamichi K. A Prefrontal Neural Circuit for Maternal Behavioural Learning in Mice. bioRxiv. 2023 Feb 4:2023-02.
[29] Donegan D, Kanzler CM, Büscher J, Viskaitis P, Bracey EF, Lambercy O, Burdakov D. Hypothalamic control of forelimb motor adaptation. Journal of Neuroscience. 2022 Aug 10;42(32):6243-57.
[30] Amengual JL, Di Bello F, Ben Hadj Hassen S, Ben Hamed S. Distractibility and impulsivity neural states are distinct from selective attention and modulate the implementation of spatial attention. Nature Communications. 2022 Aug 15;13(1):4796.
[31] Olshausen BA, Field DJ. What is the other 85 percent of V1 doing. L. van Hemmen, & T. Sejnowski (Eds.). 2006;23:182-211.

[32] Klink PC, Aubry JF, Ferrera VP, Fox AS, Froudist-Walsh S, Jarraya B, Konofagou EE, Krauzlis RJ, Messinger A, Mitchell AS, Ortiz-Rios M. Combining brain perturbation and neuroimaging in non-human primates. NeuroImage. 2021 Jul 15;235:118017.
[33] Banerjee A, Egger R, Long MA. Using focal cooling to link neural dynamics and behavior. Neuron. 2021 Aug 18;109(16):2508-18.
[34] Marr D. Vision: A computational investigation into the human representation and processing of visual information. MIT press; 2010 Jul 9.
[35] Lindsay GW, Miller KD. How biological attention mechanisms improve task performance in a large-scale visual system model. ELife. 2018 Oct 1;7:e38105.
[36] Wang X, Jadi M. Focal vs Diffuse: Mechanisms of attention mediated performance enhancement in a hierarchical model of the visual system. BioRxiv. 2023:2023-05.
[37] Fox KJ, Birman D, Gardner JL. Gain, not concomitant changes in spatial receptive field properties, improves task performance in a neural network attention model. Elife. 2023 May 15;12:e78392. **\*\* This paper manipulates activity in a convolutional neural network in order to mimic the effects of spatial attention. Importantly, they are able to independently test the behavioral impact of different hypothesized mechanisms of attention in order to identify gain changes as the most impactful.**
[38] Wimmer K, Nykamp DQ, Constantinidis C, Compte A. Bump attractor dynamics in prefrontal cortex explains behavioral precision in spatial working memory. Nature neuroscience. 2014 Mar;17(3):431-9.
[39] Laing CR, Chow CC. A spiking neuron model for binocular rivalry. Journal of computational neuroscience. 2002 Jan;12:39-53.
[40] Yamins DL, DiCarlo JJ. Using goal-driven deep learning models to understand sensory cortex. Nature neuroscience. 2016 Mar;19(3):356-65.
[41] Sajjad H, Durrani N, Dalvi F. Neuron-level interpretation of deep nlp models: A survey. Transactions of the Association for Computational Linguistics. 2022 Nov 22;10:1285-303.
[42] Dai D, Dong L, Hao Y, Sui Z, Chang B, Wei F. Knowledge neurons in pretrained transformers. arXiv preprint arXiv:2104.08696. 2021 Apr 18.
[43] Ukita J. Causal importance of low-level feature selectivity for generalization in image recognition. Neural Networks. 2020 May 1;125:185-93.
[44] Kriegeskorte N, Douglas PK. Interpreting encoding and decoding models. Current opinion in neurobiology. 2019 Apr 1;55:167-79.
[45] Buzsáki G. The brain–cognitive behavior problem: a retrospective. eneuro. 2020 Jul 1;7(4).